\documentclass[%
reprint,
superscriptaddress,
showpacs,
nofootinbib,
nobibnotes,
 amsmath,amssymb,
 aps,
prc,
floatfix,
]{revtex4-1}

\usepackage{graphicx}
\usepackage{dcolumn}
\newcolumntype{.}{D{.}{.}{-1}}
\usepackage{bm}
\usepackage{float,array,booktabs,amsmath,amssymb}

\begin{document}

\preprint{APS/123-QED}
 \title{\boldmath Isospin Mixing and the Cubic Isobaric Multiplet Mass Equation in the Lowest $T = 2,~A = 32$ Quintet
}  
\author{M.~Kamil}
\affiliation{Department of Physics and Astronomy, University of the Western Cape, P/B X17, Bellville 7535, South Africa}%
\author{S.~Triambak}
\email{striambak@uwc.ac.za}
\affiliation{Department of Physics and Astronomy, University of the Western Cape, P/B X17, Bellville 7535, South Africa}%
\author{A.~Magilligan}
   \affiliation{Department of Physics and Astronomy and National Superconducting Cyclotron Laboratory,
 Michigan State University, East Lansing, Michigan 48824-1321, USA}
\author{A.~Garc\'ia}
 \affiliation{Department of Physics and Center for Experimental Nuclear Physics and Astrophysics, University of Washington, Seattle, Washington 98195, USA}
 \author{B.\,A.~Brown}
 \affiliation{Department of Physics and Astronomy and National Superconducting Cyclotron Laboratory,
  Michigan State University, East Lansing, Michigan 48824-1321, USA}
\author{P.~Adsley}
\affiliation{School of Physics, University of the Witwatersrand, Johannesburg 2050, South Africa}%
\affiliation{iThemba LABS, P.O. Box 722, Somerset West 7129, South Africa}%
\author{V.~Bildstein}
\affiliation{Department of Physics, University of Guelph, Guelph, Ontario N1G 2W1, Canada}%
\author{C.~Burbadge}
\affiliation{Department of Physics, University of Guelph, Guelph, Ontario N1G 2W1, Canada}%
\author{A.~Diaz Varela}
\affiliation{Department of Physics, University of Guelph, Guelph, Ontario N1G 2W1, Canada}%
\author{T.~Faestermann }
\affiliation{Physik Department, Technische Universit\"{a}t M\"{u}nchen, D-85748 Garching, Germany}%
\author{P.\,E.~Garrett}
\affiliation{Department of Physics, University of Guelph, Guelph, Ontario N1G 2W1, Canada}%
\affiliation{Department of Physics and Astronomy, University of the Western Cape, P/B X17, Bellville 7535, South Africa}%
\author{R.~Hertenberger}
\affiliation{Fakult\"{a}t f\"{u}r Physik, Ludwig-Maximilians-Universit\"{a}t M\"{u}nchen, D-85748 Garching, Germany}%
\author{N.\,Y.~Kheswa}
\affiliation{iThemba LABS, P.O. Box 722, Somerset West 7129, South Africa}%
\author{K.\,G.~Leach}
\affiliation{Department of Physics, Colorado School of Mines, Golden, Colorado 80401, USA}
\author{R.~Lindsay}
\affiliation{Department of Physics and Astronomy, University of the Western Cape, P/B X17, Bellville 7535, South Africa}%
\author{D.\,J.~Mar\'in-L\'ambarri}
\affiliation{Department of Physics and Astronomy, University of the Western Cape, P/B X17, Bellville 7535, South Africa}%
\author{F.\,Ghazi~Moradi}
\affiliation{Department of Physics, University of Guelph, Guelph, Ontario N1G 2W1, Canada}%
\author{N.\,J.~Mukwevho}
\affiliation{Department of Physics and Astronomy, University of the Western Cape, P/B X17, Bellville 7535, South Africa}%
\author{R.~Neveling}
\affiliation{iThemba LABS, P.O. Box 722, Somerset West 7129, South Africa}%
\author{J.\,C.~Nzobadila~Ondze}
\affiliation{Department of Physics and Astronomy, University of the Western Cape, P/B X17, Bellville 7535, South Africa}%
\author{P.~Papka}
\affiliation{Department of Physics, Stellenbosch University, Private Bag X1, Matieland, 7602, South Africa}
\affiliation{iThemba LABS, P.O. Box 722, Somerset West 7129, South Africa}
\author{L.~Pellegri}
\affiliation{School of Physics, University of the Witwatersrand, Johannesburg 2050, South Africa}%
\affiliation{iThemba LABS, P.O. Box 722, Somerset West 7129, South Africa}%
\author{V.~Pesudo}
\affiliation{Department of Physics and Astronomy, University of the Western Cape, P/B X17, Bellville 7535, South Africa}%
\author{B.\,M.~Rebeiro}
\affiliation{Department of Physics and Astronomy, University of the Western Cape, P/B X17, Bellville 7535, South Africa}%
\author{M.~Scheck}
\affiliation{School of Computing, Engineering, and Physical Sciences, University of the West of Scotland, Paisley PA1 2BE, United Kingdom}
\author{F.\,D.~Smit}
\affiliation{iThemba LABS, P.O. Box 722, Somerset West 7129, South Africa}%
\author{H.\,-F.~Wirth}
\affiliation{Fakult\"{a}t f\"{u}r Physik, Ludwig-Maximilians-Universit\"{a}t M\"{u}nchen, D-85748 Garching, Germany}%
 
\date{\today}
%
 \begin{abstract}
The Isobaric Multiplet Mass Equation (IMME) is known to break down in the
first $T = 2, A = 32$ isospin quintet. In this work we combine high-resolution experimental data with state-of-the-art shell-model calculations to investigate isospin mixing as a possible cause for this violation. 
The experimental data are used to validate isospin-mixing matrix elements calculated with newly developed shell-model Hamiltonians. 
Our analysis shows that isospin mixing with non-analog $T = 1$ states  contributes to the IMME breakdown, making the requirement of an anomalous cubic term inevitable for the multiplet. 
\end{abstract}

\maketitle
If nuclear isospin $T$ were a conserved quantity, the members of an isobaric multiplet would be $(2T + 1)$-fold degenerate. However, it is known~\cite{Benenson:79} that this degeneracy is broken by two-body charge-dependent interactions, which can be described at tree-level as the sum of an isoscalar, isovector and isotensor operator of rank~2. To first order, the energy spacings between the multiplet members can be obtained from the expectation value of the charge-dependent perturbation. On applying the Wigner-Eckart theorem to the perturbing Hamiltonian, the mass splittings are described by the isobaric multiplet mass equation (IMME)~\cite{Wigner:57, Weinberg:59}
\begin{equation}  
 M(T_z) = a +bT_z + cT_z^2,
 \label{eq:imme} 
\end{equation} 
where each member of the multiplet is characterized by its isospin projection $T_z = (N - Z)/2$.
 
The general success of the IMME over a large mass range made it a reliable tool to address a variety of research problems. For example, it was used to test recent advances in nuclear theory~\cite{Holt:13,Ormand:17,martin:21}, map the proton dripline~\cite{Ormand:97}, identify candidates for two-proton radioactivity~\cite{Dossat:05,Blank:08}, search for physics beyond the standard model~\cite{Eric:99},
infer rapid proton capture ($rp$) nuclear reaction rates relevant for studies of novae and x-ray bursts~\cite{Wrede:09,Richter:13,Ong:17}, assess global nuclear mass model predictions~\cite{Liu:11} and constrain calculations relevant for  CKM unitarity tests~\cite{TH:15}.
%
\begin{table}[b]
\begin{flushleft}
\caption{Cubic IMME fit to measured mass excesses of the
lowest $T = 2$ quintet in $A = 32$. The fit yields $d = 0.89(11)$~keV, with $P(\chi^2,\nu) = 0.95$.}
\label{tab:IMME}
\begin{ruledtabular}
\begin{tabular}{cc..}
\multicolumn{1}{c}{Isobar}&\multicolumn{1}{c}{$T_z$}&\multicolumn{1}{c}{$M_{\rm exp}$ (keV)$^a$}& \multicolumn{1}{r}{$M_{\rm IMME}$ (keV)}\\
 \colrule
$^{32}{\rm Ar}$ & $-2$ & -2200.4(1.8) & -2200.35(158)\\
$^{32}{\rm Cl}$ & $-1$ & -8288.4(7)^b&  -8288.43(47)\\
$^{32}{\rm S}$ & $0$ &  -13967.58(28)^c  & -13967.57(25)\\
$^{32}{\rm P}$ & $+1$ & -19232.44(7)^d & -19232.43(7)\\
$^{32}{\rm Si}$ & $+2$ & -24077.69(30) & -24077.69(30)\\
\end{tabular}
$^a$ Ground state masses are taken from Ref.~\cite{Wang:21}.\\ 
$^b$ $E_x = 5046.3(4)$~keV from Ref.~\cite{Bhattacharya:08}.\\   
$^c$ $E_x = 12047.96(28)$~keV from Ref.~\cite{Triambak:06}.\\   
$^d$ $E_x = 5072.44(6)$~keV from Ref.~\cite{Endt:98}.\\    
\end{ruledtabular}
 \end{flushleft}
 \end{table}

In this context, the lowest isospin $T = 2$ quintet for $A = 32$ (with spin and parity $J^\pi = 0^+$) is an interesting case. The $\beta$ decay of $^{32}$Ar, the most proton-rich member of the quintet was previously used for searches of exotic scalar~\cite{Eric:99} and tensor~\cite{Araujo} weak interactions as well as for benchmarking isospin symmetry breaking (ISB) corrections~\cite{Bhattacharya:08} important for obtaining a precise value of $V_{ud}$, the up-down element of the CKM quark-mixing matrix~\cite{TH:15}. In fact, the $A = 32$ quintet is one of the most extensively studied and precisely measured multiplets to date~\cite{Triambak:06,ania:09,Kankainen:10,blaum:03,Signoracci:11,Lam_prc:13}. It remains an anomalous case, for which the  IMME breaks down  significantly~\cite{audi}. A satisfactory fit to the measured masses is only obtained with an additional cubic $dT_z^3$ term, with $d = 0.89(11)$~keV (c.f. Table~\ref{tab:IMME}). This is the \textit{smallest} and \textit{most precisely} determined violation of the IMME observed so far.  Unlike other multiplets, where apparent violations of the IMME were resolved 
through subsequent measurements~\cite{Herfurth,Pyle,Gallant,Glassman,Zhang:12,Su}, the $A = 32$ anomaly has persisted over several years, 
despite high-precision remeasurements of ground state masses~\cite{ania:09,Kankainen:10,Shi} as well as excitation energies~\cite{Triambak:06, Pyle}. A recent compilation~\cite{audi} showed the $A = 32$ quintet to be a unique case, in which the $\chi^2$ value for a cubic fit yields 95\% probability that it is the correct model to describe the data.   
Since there are no known fundamental reasons that preclude a cubic IMME term, it is interesting that the magnitude of the extracted $d$ coefficient for this case agrees well with 
theoretical estimates that used a simple nonperturbative model~\cite{Henley:69} or a three-body second-order Coulomb interaction~\cite{Bertsch:70}, both of which allow a non-vanishing cubic term, with $\lvert d \rvert \approx 1~{\rm keV}$. Alternatively, the role of isospin-mixing with non-analog $0^+$ states was also theoretically investigated in the recent past~\cite{Signoracci:11,Lam_prc:13}.

\begin{figure}[t]
\includegraphics[scale=0.27]{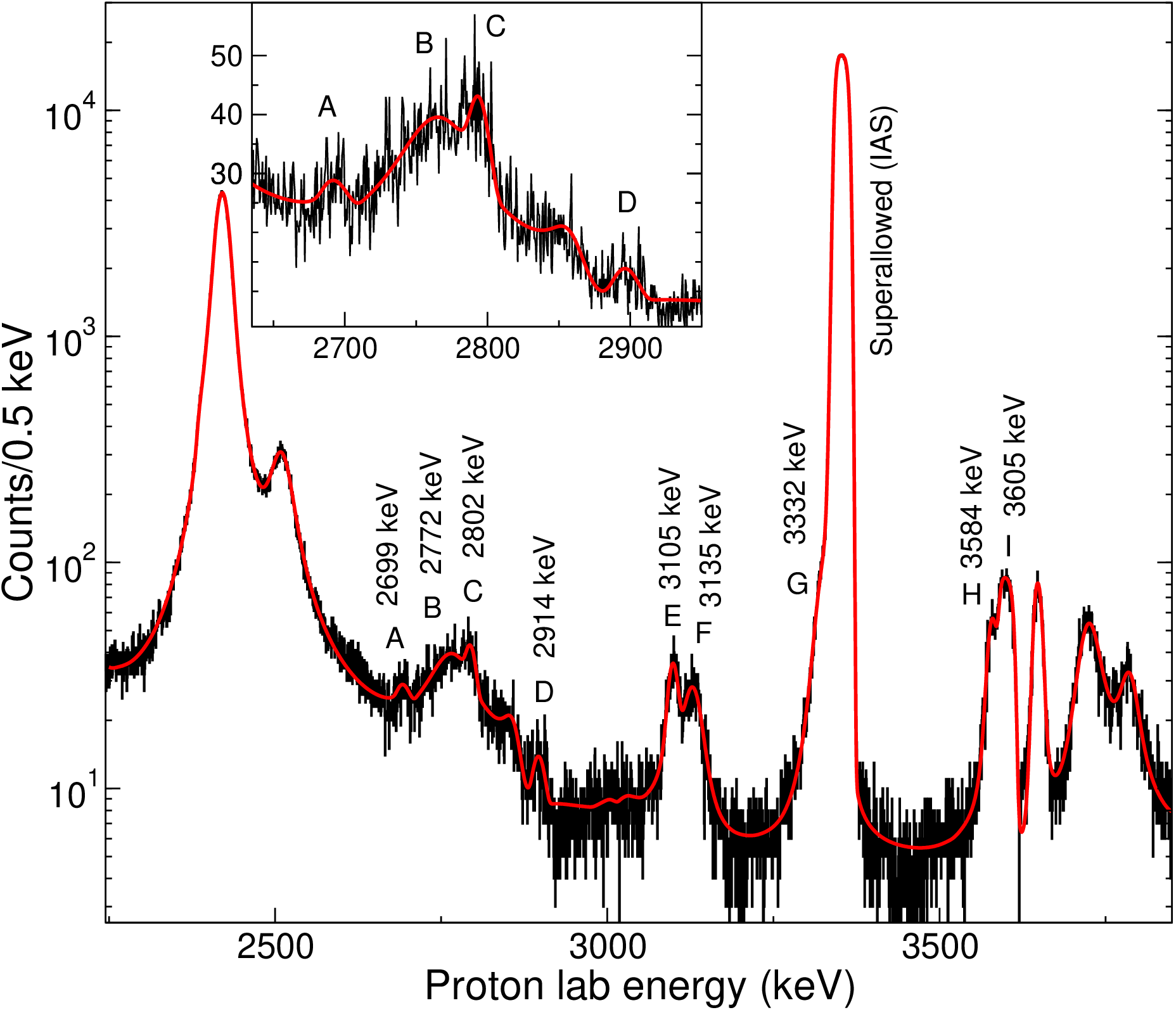}\caption{\label{fig:isolde}$^{32}$Ar $\beta$-delayed proton spectrum from the ISOLDE experiment~\cite{Eric:99} and its corresponding R-matrix fit. The inset shows a magnified portion of the spectrum.
} 
\end{figure}

We delve into the above aspect here, via an analysis of high-resolution experimental data and a comparison with calculations that use recently developed shell model Hamiltonians~\cite{usdcusdi}. For the former, we mainly rely on data from a previous $^{32}$Ar $\beta$ decay experiment at CERN-ISOLDE~\cite{Eric:99}, that acquired $\beta$-delayed protons from unbound states in the daughter $^{32}$Cl ($S_p \approx 1581$~keV) with high resolution (full widths at half maximum of $\sim 6$~keV).
%
%
The primary goal of the ISOLDE experiment was to search for scalar currents in the weak interaction, by
determining the $\beta\nu$ angular correlation ($a_{\beta\nu}$) for the decay, via a precise analysis of the shape of the superallowed $\beta$-delayed proton peak~\cite{Eric:99}. 
Part of the proton spectrum is shown in Fig.~\ref{fig:isolde}. 

The high resolution nature of the ISOLDE data allow an identification of potential isospin admixtures to the $T = 2$ isobaric analog state (IAS) in $^{32}$Cl. The nature of each $\beta$ transition is encoded in the shapes of the proton groups, which would be different if the transitions were Fermi ($0^+$ $\to$ $0^+$),  with $a_{\beta\nu} = 1$ or Gamow-Teller ($0^+$ $\to$ $1^+$), with $a_{\beta\nu} = -1/3$.  
We analyzed these data using the R-matrix formalism described in Refs.~\cite{Barker:71,Warburton}. In the analysis, the proton peaks were grouped as $p_0,~p_1,~p_2$ or $p_3$ depending on whether the proton emission left the residual $^{31}$S nucleus in its ground state or any of its first three excited states at 1249, 2234 and 3077~keV (see Fig.~9 in Ref.~\cite{Bhattacharya:08}). Interference was allowed between all levels that had the same quantum numbers, transition type (Fermi or Gamow-Teller), and final states in $^{31}$S. 
The R-matrix fits folded in the detector response function and lepton recoil effects (described in Ref.~\cite{Eric:99}), and were parameterized using various $J^\pi$ values for the daughter $^{32}$Cl states and associated $a_{\beta\nu}$ coefficients. The fits yielded relative intensities, $^{32}$Cl excitation energies and intrinsic widths. They were repeated for different values of $a_{\beta\nu}$, spin-parity combinations and $p_0,~p_1,~p_2$, $p_3$ assignments for the daughter levels to obtain best agreement with experimental data. A few important features of the analysis are described below. 
%

Peaks C, E and H were assumed to be from the $p_1$ group. These assignments were based on data reported by independent $^{32}$Ar $\beta$-delayed proton-$\gamma$ coincidence measurements~\cite{Bhattacharya:08, Blank}. We observe that a reasonably a good R-matrix fit is attained (Fig.~\ref{fig:isolde}) with the parameters listed in Table~\ref{tab:scenarios}. 
The fit assumes that peak B arises from a Fermi transition, while the others (apart from peak I) are exclusively from Gamow-Teller decays. Based purely on $\chi^2$ values from independent fits, peak I could be either from a Fermi or Gamow-Teller decay.     
\begin{table}[t]
\begin{flushleft}
\caption{R-matrix fit results for the ISOLDE data. $I_p^{\rm rel}$ is the intensity relative to the $p_0$ superallowed proton group. The last column lists corresponding states observed via the $^{32}{\rm S}(^{3}{\rm He},t)$ reaction.
}
\label{tab:scenarios}
\begin{ruledtabular}
\begin{tabular}{lcccccc}
\multicolumn{1}{c}{Peak}&\multicolumn{1}{c}{Group}&\multicolumn{4}{c}{$^{32}$Ar $\beta$ decay~\cite{Eric:99}} &\multicolumn{1}{c}{$^{32}{\rm S}(^3{\rm He},t)$}\\
  \cline{3-6}\\[-0.9em]
 & & $a_{\beta\nu}$ & $E_x$~(keV)& $\Gamma$~(keV)& $I_p^{\rm rel}$~(\%)& $E_x$~(keV)\\
 \colrule\\[-0.9em]
   A & $p_0$ & $-1/3$ &$4366(4)$ & $ < 1$ & 0.23(3)&  $4356(5)$\\
   B & $p_0$ & 1 &$4443(3)$ & $77(15)$ & 0.8(1)&  ...\\
   C& $p_1$ & $-1/3$ &5721(4) &$11(3)$  & 0.10(3)  &...\\
   D & $p_0$ & $-1/3$ & $4588(4)$ & $30(4)$ & 0.20(3)&4584(5)\\
   E & $p_1$ & $-1/3$ & $6034(2)$ & $13(3)$ &  0.14(2) & ..\\ 
   F & $p_0$ & $-1/3$ & $4817(2)$ & $26(5)$ & 0.26(3) & 4815(5)\\ 
   G & $p_0$ & $-1/3$ & $5020(2)$ & $21(2)$  & 0.49(6)&5020(5) \\    
   H & $p_1$ & $-1/3$ & $6530(2)$ & $10(3)$ & 0.25(3) &...\\
   I & $p_0$ & $-1/3$ or 1 & $5302(2)$ & $\leq 1$ & 0.45(4)  &...\\
   \cline{3-6}\\[-0.9em]
& & \multicolumn{4}{c}{$\chi^2/\nu$ = 0.80}\\
   \end{tabular}
\end{ruledtabular}
 \end{flushleft}
 \end{table}
%

%
\begin{figure}[t]
\includegraphics[scale=0.34]{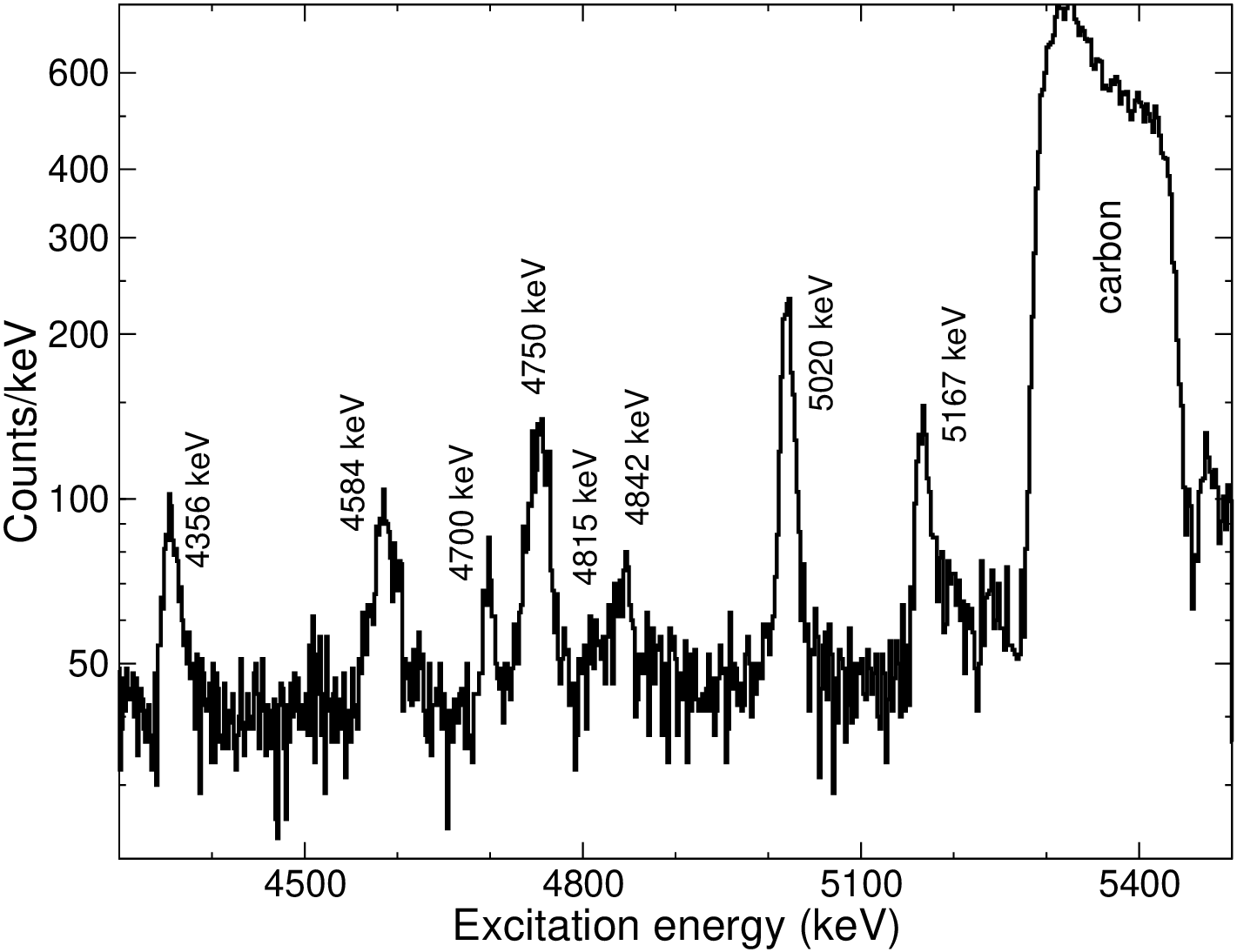}
\caption{\label{fig:spectrum}Triton spectrum from $^{32}{\rm S}(^3{\rm He},t)$ at $\theta_{\rm lab} = 10^\circ$. 
}
\end{figure}

We compared these results with $^{32}{\rm S}(^3{\rm He},t)$ data that were independently obtained at the MLL tandem accelerator facility in
Garching, Germany.
The experiment used $\sim$300~enA of 33~MeV $^{3}{\rm He}^{++}$ ions, incident on a $120~\mu$g/cm$^2$-thick natural ZnS target. The tritons exiting the target were momentum analyzed using the high-resolution Q3D magnetic spectrograph~\cite{Loffler:1973,npn_q3d}. 
A sample triton spectrum in the energy range of interest is shown in Fig.~\ref{fig:spectrum}. 
%
These data provided an important confirmation of the $p_0$ assignments for peaks A, D, F and G in our R-matrix analysis. Additionally, since the $^{32}{\rm S}(^{3}{\rm He},t)$ charge-exchange reaction predominantly populates $J^\pi = 1^+$, $T = 1$ levels at forward angles~\footnote{This assumes no anomalous isospin-mixing mechanisms within $^{32}$S.}, the states observed at these energies in \textit{both} $^{32}$Ar $\beta$ decay and the $^{32}{\rm S}(^{3}{\rm He},t)$ reaction can be ruled out as sources of $J^\pi = 0^+$ isospin impurity. This comparative analysis leaves only the 4443 and 5302~keV levels (c.f. Table~\ref{tab:scenarios}) as potential admixed states. 
We find from the $\beta$ decay data that the $p_1$ intensity for the latter is around 1.2 times larger than its $p_0$ group. In comparison, the $p_1$ intensity for the IAS is roughly 80~times smaller than the $p_0$. This is due to the low penetrability of $\l = 2$  protons from the $J^\pi = 0^+$ IAS. The above discrepancy makes it highly unlikely for the 5302~keV state to have spin-parity $0^+$, which rules it out as a source of isospin mixing.

We next used the measured $\beta$-delayed proton intensities in Table~\ref{tab:scenarios}, together with shell model calculations of isospin mixing to investigate the matter further. 
For the latter we used newly developed isospin non-conserving (INC) USDC and USDI interactions, described extensively in
Ref.~\cite{usdcusdi}.  The INC parameters in the new USD Hamiltonians were obtained from a fit to several mirror displacement energies and stringently tested via a comparison with experimental data~\cite{usdcusdi}. The isospin-mixing matrix elements calculated with these Hamiltonians were robustly validated~\cite{usdcusdi}
with results from independent high-precision  $^{31,32}$Cl $\beta$ decay experiments~\cite{s31,s32,dan}, where large isospin-mixing in the daughter $^{31,32}$S states were observed. 
More recently, such calculations were used together with a $^{32}$Ar $\beta$ decay measurement~\cite{Blank}, 
that acquired valuable proton-gamma coincidence data, albeit with lower proton energy resolution. Ref.~\cite{Blank} identified two possible sources of $T = 1$ isospin mixing at 4799 and 4561~keV. However, their measured proton branches were significantly lower than calculated values. We show below that the higher-resolution ISOLDE data justifies ruling out these proposed levels, while providing a viable alternative for the admixed $T = 1, 0^+$ state, which is consistent with both theory predictions as well as experimental observations.  

\begin{figure}[t]
    \centering
    \includegraphics[width=.45\textwidth]{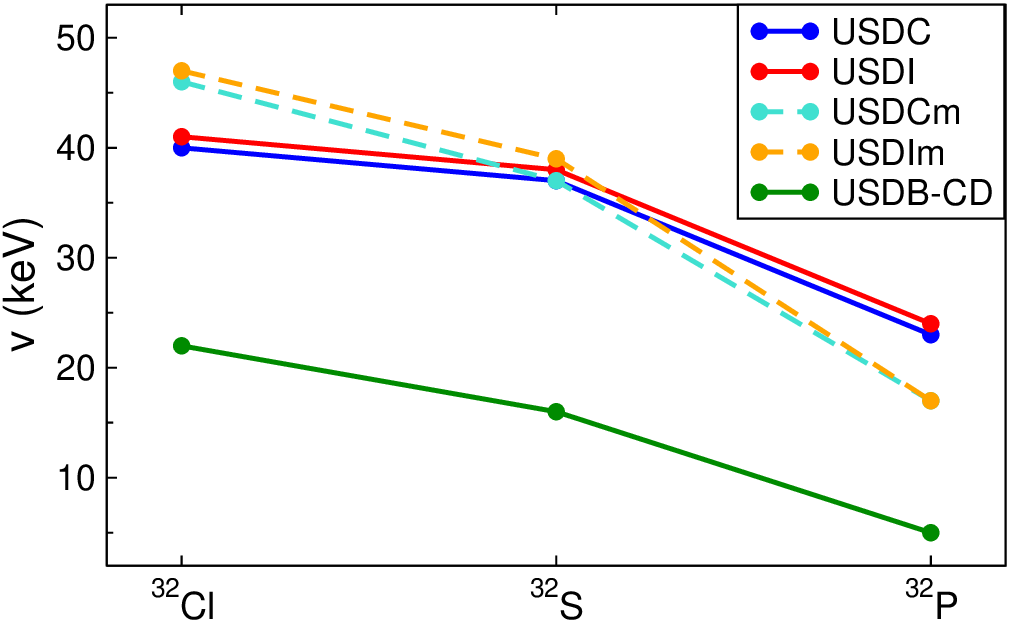}
    \caption{Evaluated isospin mixing matrix elements ($v$) using various interactions}
    \label{fig:isomixing}
\end{figure}

Our shell model calculations show that the isospin mixing within the $T_z = 1,~0,~\mathrm{and}~-1$ members of the quintet occurs primarily with a single $T=1$ state, located few hundred keV below the $T = 2$ IAS in each isobar. 
The results are summarized in Table~\ref{tab:delE}, which lists the energy differences ($\Delta E  = E_i - E_{\rm IAS}$) between the admixed $T=1$ and $T = 2$ states for each nucleus, and the calculated isospin-mixing matrix element ($v$) for $^{32}$Cl. The evaluated mixing matrix elements for each of the three nuclei are plotted in Fig.~\ref{fig:isomixing}.  We note that the mixing matrix elements obtained with the older USDB-CD interactions~\cite{ORMAND19891} are nearly a factor of two smaller than the ones obtained with the newer interactions, for all three isobars. This is consistent with previous observations for $^{31,32}$Cl $\beta$ decay~\cite{usdcusdi}.
\begin{table}[t]
    \caption{Calculated energy differences between the $T = 2$ IAS and the nearest $0^+,~T=1$ state in $^{32}$Cl, $^{32}$S, and $^{32}$P. The isospin mixing matrix element in $^{32}$Cl is listed for comparison.}
    \centering
    \begin{ruledtabular}
    \begin{tabular}{lcccc}
    Interaction & \multicolumn{3}{c}{$\Delta E$~(keV)} & \multicolumn{1}{c}{$v$~(keV)}\\
     \cline{2-4}\\[-0.9em]
     & \multicolumn{1}{c}{$^{32}$Cl} & \multicolumn{1}{c}{$^{32}$S} &  \multicolumn{1}{c}{$^{32}$P} &  \multicolumn{1}{c}{$^{32}$Cl} \\
     \colrule\\[-0.9em]
          USDC &     -226 &     -186 &     -237 &     40\\
     USDI &     -308 &     -266 &     -326 &     41\\
    USDCm &     -324 &     -239 &     -293 &     46\\
    USDIm &     -405 &     -321 &     -383 &     47 \\
     USDB-CD &     -440 &     -378 &     -427 &     22 \\
     \colrule\\[-0.9em]
     Expt (this work)&     -603 &   &     &    $39.0(24)$\\
    \end{tabular}
    \end{ruledtabular}
    \label{tab:delE}
\end{table}

The predicted $J^\pi;T = 0^+;1$ level in $^{32}$Cl can be identified by obtaining an experimental value of $v$ from the data in Fig.~\ref{fig:isolde} and Table~\ref{tab:scenarios}. 
For two-state mixing, $v_{\rm expt}$ is simply
\begin{equation}
v_{\rm expt} = \Delta E_{\rm expt}\left[\frac{B(F)_{\rm admix}}{B(F)_{\rm SA}}\right]^{1/2},
\label{eq:vexp}
\end{equation}
where the ratio in the square bracket is the (Fermi) strength to the admixed $T = 1$ state, relative to the superallowed decay. This is easily determined from the measured $I_p^{\rm rel}$ values in Table~\ref{tab:scenarios}, the ratio of calculated phase-space factors, a small ISB correction~\cite{Bhattacharya:08} and the $p_0$ contribution to the total superallowed intensity. On applying this prescription to the only candidate $0^+$ level at 4443~keV, we obtain a $v_{\rm expt} = 39.0(24)$~keV, in excellent agreement with the calculations. The results in Table~\ref{tab:delE}, together with our aforementioned observations and the experimental values listed in Table~\ref{tab:scenarios} 
allow a credible
identification of the 4443~keV level as the predicted admixed $T = 1$ state.
The discrepancy between theory and experiment for $\Delta E$ should not be surprising, given the $\sim$150~keV root-mean-square (rms) deviation for energies in USD interactions~\cite{usdcusdi}. 

\begin{figure}[t]
    \centering
    \includegraphics[width=.45\textwidth]{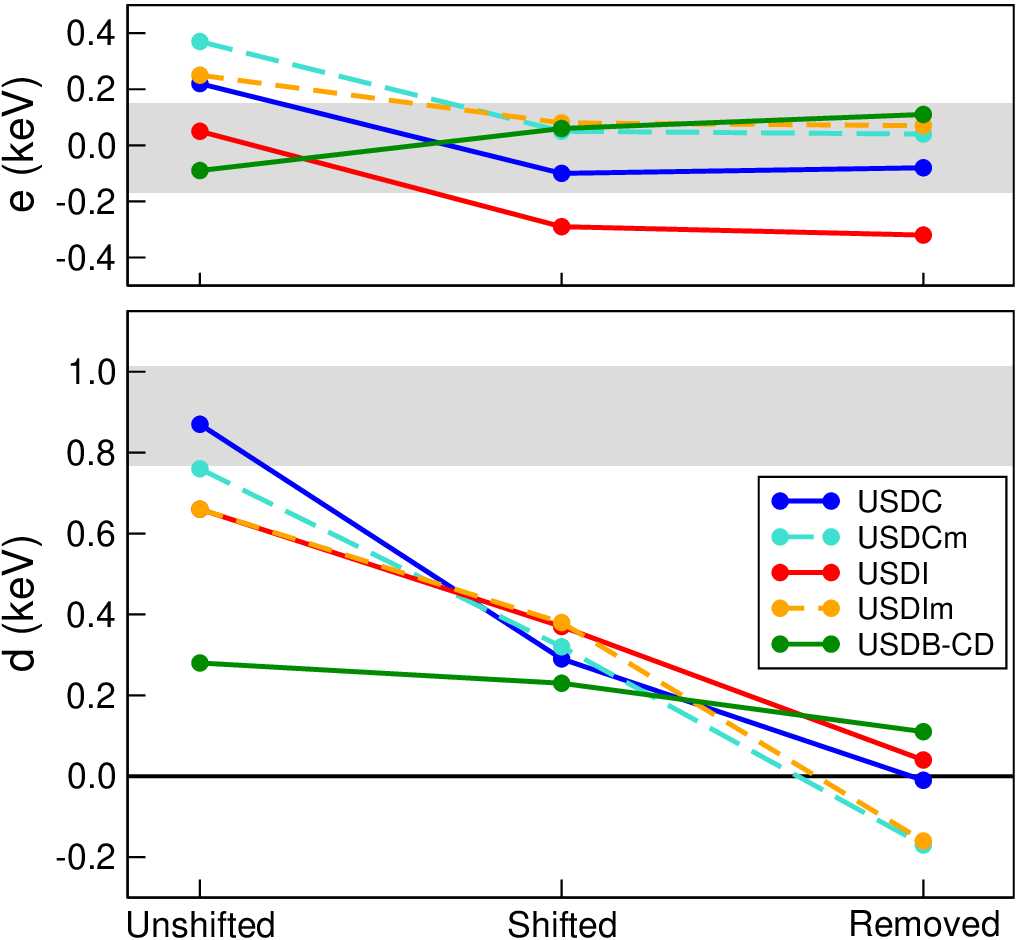}
    \caption{Extracted cubic and quartic coefficients. The three groups of results are obtained A) at face value, B) by shifting the energies of the $T = 2$ states in $^{32}$Cl, $^{32}$S and $^{32}$P to match the 603~keV energy difference observed in $^{32}$Cl, and C) on removing the $T = 1$ isospin mixing. The shaded areas correspond to experimental values.
    }  
    \label{fig:coefficients}
\end{figure}

We next investigated additional cubic ($dT_z^3$) and quartic ($eT_z^4$) terms to the IMME due to such isospin mixing. One can determine the \textit{exact} solutions for the $d$ and $e$ coefficients by modifying Eq.~\eqref{eq:imme} to incorporate such terms, such that  
\begin{equation}
\begin{aligned}
    d &= \frac{1}{12}\left( M_{2} - 2 M_{1} + 2M_{-1} - M_{-2} \right)~{\rm and} \\
    e &=  \frac{1}{24}\left( M_2 - 4 M_1 + 6 M_0 -  4 M_{-1} + M_{-2} \right), \\
\end{aligned}
\end{equation}
where the $M_{T_Z}$ are isobar masses in the quintet. The results for $d$ and $e$ using the calculated values of $v$ and $\Delta E$ are shown in Fig.~\ref{fig:coefficients}, and labeled as ``unshifted''. We repeated these evaluations by shifting the $T=2$ states in $^{32}$Cl, $^{32}$S and $^{32}$P by the amount needed to reproduce our experimentally determined 603~keV energy difference in $^{32}$Cl. The same $\Delta E$ was used for the three isobars due to the lack of similar experimental information for $^{32}$S and $^{32}$P. The shifts were accomplished by adding a $T^2$ term to the Hamiltonian that shifts the $T=2$ states relative to the others, without changing the isospin-mixing. As evident in Fig.~\ref{fig:coefficients}, the shifts mildly affect the $e$ coefficient (due to changes in the $T = 0$ mixing with the IAS in $^{32}$S), but significantly decrease the calculated $d$ coefficient to $\approx 0.3$--$0.4$~keV for the new interactions. 
%
The single-state contributions from $T = 0$ and $T = 1$ levels are 
\begin{equation}
\label{eq:coeffs}
\begin{aligned}
    d_i &= -\frac{1}{6} s_{\rm P} + \frac{1}{6} s_{\rm Cl} \newline \\
    e_i &= -\frac{1}{6} s_{\rm P} + \frac{1}{4} s_{\rm S} - \frac{1}{6} s_{\rm Cl}, 
\end{aligned}
\end{equation}
where $s = -v^2/\Delta E$ is the shift in each IAS due to two-state mixing. Thus, one can remove the $T = 1$ mixing contribution for further investigation (labeled as ``removed'' in Fig.~\ref{fig:coefficients}). 
We observe that on doing so, the extracted coefficients are mostly consistent with zero. The negative $e$ coefficient from the USDI calculation is due to mixing with a $T = 0$ state in $^{32}$S. However such $T = 0$ mixing would not explain the non-zero $d$ coefficient required for the quintet, as evident from Eq.~\eqref{eq:coeffs}. 

The above analysis validates the contention that isospin mixing with predicted $T = 1$ levels \textit{necessitates} a small cubic term for the multiplet. 
Our extracted $d$ coefficients for the ``shifted'' calculations from different USDC and USDI Hamiltonians agree reasonably well with one another, but are smaller than the experimental value $d = 0.89(11)$~keV, from Table~\ref{tab:IMME}. 
\begin{table}[t]
\begin{flushleft}
    \caption{Calculated proton emission amplitudes from states in $^{32}$Cl, compared with experiment. The last column lists calculated isospin mixing corrections for $^{32}$Ar superallowed Fermi decay.}
        \label{tab:widths}
    \begin{ruledtabular}
    \begin{tabular}{llllc}
    \multicolumn{1}{l}{Interaction} & \multicolumn{3}{c}{Proton emission amplitudes $(A)$}&\multicolumn{1}{c}{$\delta_{C}^{\rm cm}$} \\
     \cline{2-4}\\[-0.9em]
   \multicolumn{1}{l}{(shifted calculation)}  & \multicolumn{1}{l}{$T=2$} & \multicolumn{1}{l}{$T=2$} &  \multicolumn{1}{l}{$T=1$} & (\%)\\
     & \multicolumn{1}{l}{$(p_0)$} & \multicolumn{1}{l}{$(p_1)$} &  \multicolumn{1}{l}{$(p_0)$} &  \\
    \colrule\\[-0.9em]
     USDC &     0.011 &     0.022 &     0.21 &  0.55\\
     USDI &     0.011&      0.031 &     0.19  & 0.58\\
    USDCm &     0.0052 &     0.031 &     0.21  & 0.15  \\
    USDIm &     0.0043 &     0.031 &     0.19       & 0.70   \\
     USDB-CD &   0.0024 &     0.017 &     0.21      & 0.15  \\
    \colrule\\[-0.9em]
    $\Gamma_{\rm sp}$~(keV) &990&17.5&590\\
    $\Gamma_{\rm expt}$~(keV) & 0.0182(5)$^a$&0.000233(7)$^a$&77(15)$^b$ &\\
     $A_{\rm expt}$&     0.0041(1)$^a$ &  0.0035(1)$^a$  &   0.34(4)$^b$ &  \\
    \end{tabular}
    \end{ruledtabular}
     $^a$ From Ref.~\cite{Bhattacharya:08}.\\
    $^b$ This work, using data from Ref.~\cite{Eric:99}. 
    \end{flushleft}
\end{table}

As further tests of our calculations, we also evaluated amplitudes for isospin-forbidden proton emission from the two admixed $J^\pi = 0^+$ levels in $^{32}$Cl and the effect of the $T = 1$ isospin mixing on the superallowed Fermi decay of $^{32}$Ar. Unlike the energy shift of the $T = 2$ IAS in $^{32}$Cl, which is predominantly from isospin mixing with the predicted 0$^{ + }_{2}$ $T = 1$ state below the IAS, isospin-forbidden proton emission from the IAS depends on $T = 1$ mixing with a large number of states in $^{32}$Cl and isospin mixing within $^{31}$S, which is dominated by mixing of the lowest $T = 3/2$ state into its ground state.

We calculated proton widths for $p_0$ and $p_1$ transitions from the $0_2^+$ admixed $T = 1$ state and the $T = 2$ IAS in $^{32}$Cl. The widths were evaluated using the simple expression
\begin{equation}
\Gamma _{\rm th} = (C^2S)(32/31)^{2} \Gamma _{\rm sp},
\end{equation}
where the $(32/31)^2$ factor is a center-of-mass correction~\cite{cm}, the
$C^2S$ are the shell model spectroscopic factors, and $\Gamma_{\rm sp}$ are single-particle proton widths. Similar to Ref.~\cite{Signoracci:11}, the $\Gamma_{\rm sp}$ were calculated from $p$ + $^{31}{\rm S}$ scattering on potentials obtained with an energy-density functional calculation with the Skx Skyrme-type interaction~\cite{skx}. On the other hand, the measured proton widths~\cite{Bhattacharya:08} for the $T = 2$ IAS are known to be $18.2(5)$~eV and $0.233(7)$~eV for the $p_0$ and $p_1$ protons respectively. Together with the calculated single-particle proton decay widths, we use these results to obtain experimental values for the decay amplitudes $  A  = (C^2S)^{1/2} $. These are simply determined from the relation $\Gamma _{\rm expt} = A_{\rm expt}^{2}  (32/31)^{2} \Gamma _{\rm sp}$. The results for $A_{\rm expt}$ are shown in Table~\ref{tab:widths} and compared with theory predictions, obtained using the `shifted' calculations.\footnote{For example, for the shifted USDCm calculation, the calculated amplitude can be decomposed as $  A  $ = 0.0041 (0$^{ + }_{1}$; $T=1$) $+$ 0.0159 (0$^{ + }_{2}$; $T=1$)
$+$ 0.0009 (0$^{ + }_{3}$; $T=1$) $+$ 0.0009 (all other $T=1$) $-0.0166$ ($^{31}$S; $T=3/2$)
= 0.0052. The USDC result has a larger amplitude, mainly due to a 50\% smaller destructive contribution from the $T=3/2$ state in $^{31}$S.} We observe reasonable agreement between theory and experiment, except for the $T = 2$ $p_1$ transition, whose calculated amplitudes are found to be much larger. 

Finally, we also provide isospin-symmetry-breaking (ISB) corrections for $^{32}$Ar $\to$ $^{32}$Cl superallowed Fermi decay, due to the isospin-mixing in $^{32}$Cl. The $T = 2$ $\to$ $T = 2$ superallowed strength is reduced by a factor $(1 - \delta_C)$, where $\delta_C$ is the total ISB correction~\cite{TH:15}. Such corrections play a critical role in testing the unitarity of the CKM matrix and placing important constraints on beyond the standard model (BSM) physics~\cite{TH:15}. The ISB correction is generally expressed as a sum of two separate contributions, $\delta_C = \delta_C^{\rm cm} + \delta_C^{\rm ro}$~\cite{Bhattacharya:08}, from configuration mixing and a overlap mismatch between the parent and daughter radial wavefunctions. The former are known to quite model dependent as they are very sensitive to the details of their calculation~\cite{TH:15}. Our calculated results for $\delta_{C}^{\rm cm}$ (from the $T = 1$ mixing in $^{32}$Cl) are listed the final column of Table~\ref{tab:widths}. It may be noted that for the \textit{shifted} USDCm and USDIm calculations, which show best agreement with the measured $T = 2$ $p_0$ amplitude, we obtain $\delta_{C}^{\rm cm}$ = 0.15\% and 0.70\% respectively.  From a previous evaluation of $\delta_C^{\rm ro} = 1.4\%$~\cite{Bhattacharya:08}, these yield $\delta_C = 1.6\%$ and $2.1\%$, in agreement with the experimentally extracted value, $\delta_C^{\rm expt} = 2.1(8)\%$~\cite{Bhattacharya:08}.

In summary, we used high-resolution experimental data to validate newly-developed shell model calculations of isospin mixing in $^{32}$Cl. This analysis is used to investigate the observed IMME violation in the first $T = 2,~A = 32$ quintet. We show that isospin mixing with shell-model-predicted $T = 1$ states below the IAS necessarily result in a break down of the IMME, leading to the requirement of a small cubic term.  However, this alone cannot explain the magnitude of the experimental $d$ coefficient in Table~\ref{tab:IMME}.  Experimental investigations of intruder $0^+$ levels, isospin-mixing in $^{32}$S and $^{32}$P, continuum coupling of the proton unbound states in $^{32}$Cl, and further mass measurements may be useful in this regard.    

Our observations pertaining to $^{32}$Ar $\to$ $^{32}$Cl superallowed Fermi decay may also be useful to benchmark theory calculations~\cite{Bhattacharya:08} of model-dependent ISB corrections that are important for top-row CKM unitarity tests~\cite{TH:15}. 
This is particularly relevant in light of recent evaluations of radiative corrections~\cite{box} that show an apparent violation of CKM unitarity at the $> 3\sigma$ level~\cite{Seng}.   
%

\begin{acknowledgments}
We thank Eric Adelberger and Gordon Ball for insightful and illuminating discussions. This work was partially supported by the National Research Foundation (NRF), South Africa under Grant No.~85100, the U.S. National Science Foundation under Grant No. PHY-1811855 and the U.S. Department of Energy under Grants No. DE-SC0017649 and DE-FG02-93ER40789.
P.A. acknowledges funding from the Claude Leon Foundation in the form of a postdoctoral fellowship. 
\end{acknowledgments}

\bibliography{imme_revised_november}

\end{document}